\documentclass[pra,aps,preprintnumbers]{revtex4}
\usepackage{epsfig,amsmath,amssymb,amsfonts}
\usepackage{subfigure}
\def\CC{{\rm\kern.24em \vrule width.04em height1.46ex depth-.07ex
\kern-.30em C}}
\def\P{{\rm I\kern-.25em P}}
\def\RR{{\rm
         \vrule width.04em height1.58ex depth-.0ex
         \kern-.04em R}}
\def\RR{{\rm\kern.24em \vrule width.04em height1.46ex depth-.07ex
\kern-.30em R}}
\def\P{{\rm I\kern-.25em P}}
\def\RR{{\rm
         \vrule width.04em height1.58ex depth-.0ex
         \kern-.04em R}}

\newcommand{\ket}[1]{\left | \, #1 \right\rangle}

\newcommand{\be}{\begin{equation}}
\newcommand{\ee}{\end{equation}}
\newcommand{\bq}{\begin{eqnarray}}
\newcommand{\eq}{\end{eqnarray}}

\begin{document}

\title{Quantum computation in optical lattices via global laser addressing}

\author{Alastair Kay}
\author{Jiannis K. Pachos}
\affiliation{Department of Applied Mathematics and Theoretical
Physics, University of Cambridge, Cambridge CB3 0WA, UK.}

\date{\today}

\begin{abstract}
A scheme for globally addressing a quantum computer is presented along with its
realisation in an optical lattice setup of one, two or three
dimensions. The required resources are mainly those necessary for
performing quantum simulations of spin systems with optical lattices,
circumventing the necessity for single qubit addressing. We present the
control procedures, in terms of laser manipulations, required to
realise universal quantum computation.
Error avoidance with the help of the quantum Zeno
effect is presented and a scheme for globally addressed error
correction is given. The latter does not require measurements during the
computation, facilitating its experimental
implementation. As an illustrative example, the pulse sequence for the
 factorisation of the number fifteen is given.

\end{abstract}

\maketitle
%\pacs{PACS numbers: 03.75.Lm, 03.67.Lx, 42.50.-p}

\section{Introduction}

%In this paragraph I have added a comment to clarify our use of the
%term `global addressing' in response to the first remark of referee
%1.
Over the past few years, Benjamin \cite{Benjamin:2000,Benjamin:2001b,
Benjamin:2002a, Benjamin:2003b}, in particular, has followed up an
initial proposal by Lloyd \cite{LLoyd:1993} on the concept of global
control schemes for quantum computation. The motivation for such
schemes is simple - instead of needing individual elements for the
manipulation of every single qubit in a system, which is
technologically very difficult, we control a very limited set of
fields that are applied to all the qubits in the system. 
It is in this sense that we use the term `global addressing' - we will
use lasers with a beam width that addresses the
whole ensemble of atoms. While every qubit will be given the same commands,
we shall demonstrate the techniques that localise the effects so as to carry
out operations on only single qubits.

Recently
\cite{Zoller:1}, a practical scheme has been proposed to implement
such ideas with optical lattices (see also \cite{Vollbrecht}). Here we
present another possible scheme, which has
a number of advantages in terms of ease of implementation and
especially scalability, where we can perform
computation in two, or even three, dimensions of an optical lattice.
As we shall see in the following, the present scheme gives the
possibility to incorporate error
correction and fault tolerance in a straightforward manner.
%nodding in the direction of where we're dealing with the referees
%comments. NO YOU HAVEN'T!!!

%added clarification as to the nature of the pointer
The way that we implement global control is relatively simple. We have
a register array of qubits on which we wish to perform computation and
an auxiliary array for which we retain single qubit control over a
certain lattice site \cite{comment}. This enables us to initialise a
pointer (referred 
to as a control unit or marker atom in previous works) which is
essentially a unique component which we can use to localise operations
while applying global operations. Initially, the pointer is just the
same atom as all other qubits, trapped in the lattice in the state
$\ket{0}$, except that we individually rotate it to the state $\ket{1}$. 
By exclusively using global
addressing, we can move this pointer atom relative to the computational
qubits and apply controlled-$U$ operations which then effects $U$ solely on the qubit adjacent to the pointer. Two-qubit gates
can be implemented in a similar way, using a three qubit gate such that, in the presence of the pointer, the desired gate is enacted on two neighbouring qubits. This control set is sufficient
to give universal quantum computation. In comparison the schemes
requiring individual addressing, the only additional resources are those
required to move the pointer around the lattice structure. For a
$d$-dimensional structure of $N$ qubits, this will increase the total
number of steps by a factor of order $\sqrt[d]{N}$.
Significantly, error avoiding and error correcting
techniques can be implemented that respect the global addressing
requirement and render our computational scheme favourable for
scalable quantum computation.

\section{Atomic system and superlattices}
\label{atomic}

\subsection{Atomic system and optical lattices}

An important tool in the manipulation of atomic ensembles is the
employment of optical lattices. These can generate one, two or
three dimensional structures of potential minima that can be used to trap atoms. In
particular, consider two species of atoms, namely $\sigma=
\{a,b\}$, that are confined within sinusoidal configurations of
optical lattices produced by lasers of wavelength $\lambda$. These
species can be two different 
internal ground states of the same atom, which can then be trapped and manipulated by two optical lattices of different polarisations. When the atoms
are cold enough and the amplitudes of the optical lattices are
sufficiently large, the atoms become restricted to the lowest
Bloch-band. Hence, the evolution of the system can be described by the
Bose-Hubbard Hamiltonian,
\begin{equation}
H=-\sum_{i\sigma} J^\sigma_i (a_{i\sigma}^\dagger
a_{i+1 \sigma} + a_{i\sigma}
a_{i+1 \sigma}^\dagger)+\frac{1}{2} \sum _{i
\sigma \sigma'} U_{\sigma \sigma'}
a^{\dagger}_{i\sigma}a^{\dagger}_{i\sigma'}a_{i\sigma'}a_{i\sigma}.
\end{equation}
This is comprised of tunnelling transitions of
atoms between neighbouring sites of the lattice, 
\begin{equation} 
J ={E_R \over 2} \exp (- {\pi^2 \over 4} \sqrt{ V_0 \over E_R})
  \left[\sqrt{V_0 \over E_R} + \left( {V_0 \over E_R}
  \right)^{3/2}\right] 
\label{tunnel}
\end{equation}
and collisional
interactions between atoms in the same site
\begin{equation} 
U={4 a_s  \over \lambda } V_0^{3/4} E_R^{1/4}
\label{collision}
\end{equation}
where $E_R=\hbar^2 k^2/(2m)$ is the recoil energy,
$k=2\pi/\lambda$, $m$ is the mass of the atoms, $V_0$ is the potential
barrier between neighbouring lattice sites and $a_s$ is the s-wave scattering length of the colliding atoms.

The collisional couplings can be
arranged to take significantly large values via
Feshbach resonances \cite{Inouye,Donley}. Small tunnelling couplings, with
respect to the collisional ones, can be produced by increasing the
amplitude of the laser fields comprising the optical lattice. In this
way, the system can be brought into the Mott insulator phase with only
one atom per lattice site \cite{Kastberg, Raithel, Greiner:2002a, Greiner:2002b}.

\begin{figure}[!htp]
\begin{center}
\includegraphics[width=12cm]{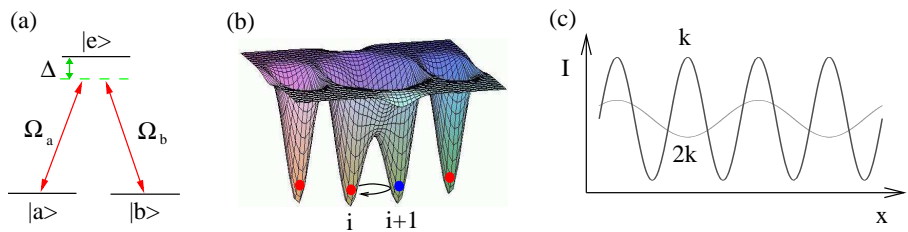}
\end{center}
\caption[comb1]{(a) The atomic levels with two ground states
coupled via an excited state by a Raman transition resulting in a
coupling of $J^R=\frac{\Omega_a\Omega_b^*}{\Delta}$.
(b) The interaction between two neighbouring sites where tunnelling is activated
by lowering the potential barrier between them.
(c) The intensity profile of employed standing waves that comprises a
superlattice, activating (b) between alternate sites.}\label{comb1}
\end{figure}

We encode the logical $|0\rangle$ and $|1\rangle$ of the computation in the
$|a\rangle$ and $|b\rangle$ ground states of the atom, which correspond
to the populations of the different modes of the optical lattice. As
seen in Figure~\ref{comb1}(a), the state of an atom can be transported between modes
$a$ and $b$ by performing Raman
transitions.

\subsection{Simulation of spin Hamiltonian}

We initially consider that the optical lattice system is brought into
the Mott insulator regime where there is only one atom per lattice
site. By manipulating the tunnelling couplings, it is possible to obtain
a nontrivial evolution suitable for performing quantum computation. It
is possible to expand the evolution due to the Bose-Hubbard Hamiltonian in
terms of the small parameters $J^\sigma/U_{\sigma \sigma'}$. Up to second order in perturbation theory, this expansion is given, in terms of the Pauli matrices
\cite{Kukl,Duan,Pachos:2004b}, by
\begin{equation}
    H = \sum_{i=1}^3 \Big[ \vec{B} \cdot \vec{\sigma}_i
    +\lambda^{(1)} \sigma^z_i \sigma^z_{i+1} + \lambda^{(2)}
    (\sigma^x_i \sigma^x_{i+1} +\sigma^y_i \sigma^y_{i+1})  \Big]. \label{ham1}
\end{equation}
The couplings $\lambda^{(i)}$ are given by
\begin{equation}
\lambda^{(1)} = {{J^a}^2+{J^b}^2\over 2 U_{ab}} - {{J^a}^2 \over
 U_{aa}}- {{J^b}^2 \over U_{bb}} \,\,,\,\,\,
\lambda^{(2)}=- {J^a J^b \over U_{ab}}  .
\nonumber
\end{equation}

Single particle phase rotations of the form $B_z
\sum_i\sigma^z_i$ can be cancelled with
\begin{displaymath}
B_z = -{2{J^a}^2\over U_{aa}}+{2{J^b}^2\over U_{bb}}.
\end{displaymath}
The local field $\vec{B}$ can then be
arbitrarily tuned by applying appropriately detuned laser fields.

The effective couplings $\lambda^{(i)}$ can be tuned by
manipulating the amplitudes of the laser fields that generate the
optical lattices. In particular, by
activating only one of the two tunnelling couplings, e.g. $J^a$, we can obtain the
diagonal interaction $\sigma^z_i \sigma^z_{i+1}$ along all the qubits
of the lattice. This, up to local qubit rotations, is equivalent to a
series of control phase gates (CP). However, if we activate both of the
tunnelling couplings with appropriate magnitudes, it is possible to
activate the exchange interaction $\sigma^x_i \sigma^x_{i+1}
+\sigma^y_i \sigma^y_{i+1}$. When applied for a sufficient time
interval it results in a SWAP gate, exchanging the atoms at
neighbouring lattices sites.
%Adding something to answer referee 2, main point about effective Hamiltonian

We have used an effective Hamiltonian to create the gates that
we are interested in. The creation of these gates is studied in more
detail in \cite{Pachos:2003a}, where the error introduced by the real
Hamiltonian is examined for experimentally sensible
parameters. Currently achievable errors are shown to be of the order
of $10^{-3}$, which is small enough for an in--principle
demonstration, even if it is not small compared to thresholds for
fault tolerance \cite{Steane:98}. Ref. \cite{Pachos:2003a} also
shows how to create such gates without resorting to an effective
Hamiltonian in the adiabatic regime, giving much shorter time scales for
gate implementation.

\subsection{Superlattices}

As we have seen in the previous section, it is possible to control the
tunnelling coupling constants by
modifying the amplitude of the standing laser fields. The way to avoid
single atom addressing is to employ
``superlattices'' (see Figure \ref{comb1}(c)), that is the superposition of optical lattices with
different wavelengths. This will eventually be
sufficient for performing universal quantum computation. With
superlattices, we can manipulate the tunnelling couplings, $J_i^\sigma$,
and consequently the effective couplings $\lambda$, by varying the
potential barrier $V_0$, as seen in equation (\ref{tunnel}). In
particular, we shall employ two independent
lattices whose spatial periods differ by a factor of two. This can be
achieved, for example, with two pairs of laser beams, each one creating
a lattice with period $d_i=\lambda/[2 \sin(\theta_i/2)]$ that depends
on the angle $\theta_i$ between them \cite{Peil,Zoller:1}. Hence,
by choosing $\theta_i$s appropriately, one can achieve a light-shift
potential given by
\begin{equation}
U(x)=U_1\cos (2 k x) +U_2 \cos (k x -\phi)
\end{equation}
where $k=2\pi\sin(\theta_1/2)/\lambda=4\pi\sin(\theta_2/2)/\lambda$
and $\phi$ is the phase difference between the second pair of
lasers, while the first pair is taken to be in phase. 
By changing the amplitudes $U_i$ and the phase $\phi$ it is possible
to obtain the control procedures necessary for the realisation of the
presented scheme. In the same way, one can create the desired three
dimensional lattice structures.

Raman transitions can be performed on every other qubit in a desired
direction by employing similar structures of standing
waves that are properly tuned, creating a two photon transition
between the atomic ground states $a$ and $b$. If we denote the laser Rabi frequencies that couple the states $a$ and $b$ by $\Omega_a$ and $\Omega_b$ respectively, then the coupling of the two states is given by $\Omega_a\Omega_b^*/\Delta$. The excited state has a detuning of $\Delta$, as shown in Figure \ref{comb1}(a).
By positioning the lasers such that the
$\Omega_\sigma$s have a sinusoidal configuration, we can activate the
Raman transition only on alternate rows. We will commonly tune this to activate the
transition on the register arrays without affecting the auxiliary
ones.

\section{Quantum computation with superlattices}

In what follows, we show how to perform one and two qubit
gates between any qubits. To do this, we need to transport the pointer
qubit to any desired location. In particular, for a single qubit gate
we first transport the pointer next to the targeted qubit, $q_1$. A
conditional unitary transformation is then applied between the auxiliary
array and the register one. This acts if the auxiliary qubit is in
the state $\ket{1}$, resulting in a gate only on the
qubit conditioned by the pointer. In order to perform a two qubit gate,
we need to perform interactions between all three qubits, the two
targeted ones, $q_1$ and $q_2$, and the pointer. 

\begin{figure}[!htp]
\begin{center}
\includegraphics[width=11.3cm]{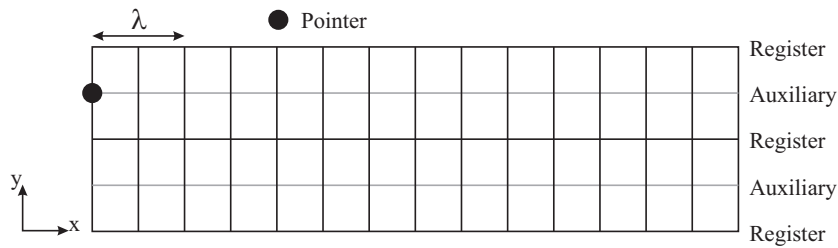}
\end{center}
\caption[lattice]{The two dimensional lattice comprising of
squares. The qubits are placed at the vertices. Computation is
performed at the register arrays, while the auxiliary arrays are used
for transporting the pointer.}\label{lattice}
\end{figure}

In principle, there is much freedom in choosing the geometry for
performing such steps. The minimal one dimensional case is rather
cumbersome. Alternatively, one can use semi-one dimensional models,
e.g. ladders, consisting of two parallel interacting arrays of qubits,
one being the auxiliary array and the other the register. In terms of control procedures, the square configuration adopted here (see
Figure \ref{lattice}) requires the least resources. The lattice comprises standing laser
fields with wavelength $\lambda$ which we assume are
perpendicularly oriented. We prepare the qubits such that one is
placed at each site of the lattice, all in the state $|0\rangle$. The pointer is then created by performing $\sigma_x$ on one of these qubits. We consider that the
register qubits are in arrays along the $x$ direction and occupy
every other lattice array along the $y$ direction. At the same time,
the auxiliary arrays between them are used for the transportation of
the pointer qubit.
The idea is to use global laser addressing to perform all the
necessary manipulations on the lattice to result in quantum
computation on the register qubits. These manipulations, with
their equivalent physical implementation, are described in the
following.

\subsection{Transport of pointer qubit along the same or different
auxiliary arrays} 

\begin{figure}[!htp]
\begin{center}
\subfigure{\includegraphics[width=13cm]{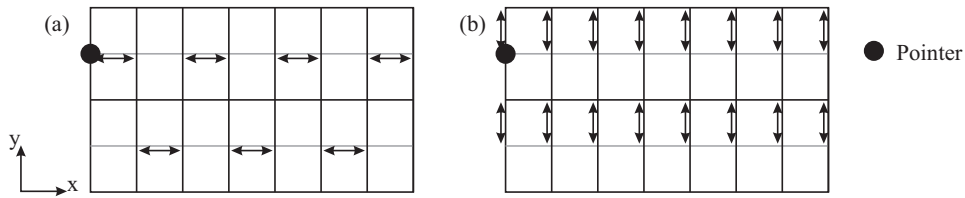}}
\end{center}
\vspace{-0.5cm}
\caption[first]{The superlattice for the swapping of the pointer
  qubit. This causes adjacent qubits to move in opposite
  directions. (a) gives the horizontal motion of the pointer qubit,
  while (b) gives the vertical motion of the auxiliary array.}\label{first}
\end{figure}

The transport of the pointer qubit along an
array can be performed by activating a lattice as given in Figure
\ref{first}(a).
The superimposed lattices create minima between the pointer qubit and
its appropriate neighbour on its left or right, resulting in
the SWAP interaction between them. By exchanging
the minima between neighbouring links, it is possible to move the
pointer to any lattice site along the $x$ direction.
To transport the pointer qubit to a different auxiliary array we have
to superimpose the optical lattice given in Figure \ref{first}(b). 
This exchanges the auxiliary and the register arrays, resulting
in the transport of the pointer along the $y$ direction. These
techniques can be combined to transport the pointer to any
location we want in order to perform a one qubit gate on the desired qubit.
This manipulation
also plays a significant role when performing two-qubit
gates. If we implement this interaction on the register arrays, then
qubits that are separated by an even number of lattice sites can be moved
next to each other. Qubits that are separated by an odd number of
sites do not move relative to each other with this swapping
mechanism.

\subsection{One-qubit gates with common addressing}

\begin{figure}[!htp]
\begin{center}
\includegraphics[width=7.64cm]{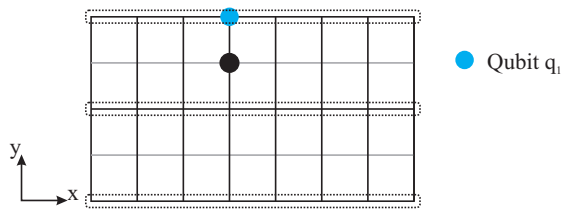}
\end{center}
\vspace{-0.5cm}
\caption[second]{The Raman transition is activated by standing
  waves along the $y$ direction that rotate only the register
  qubits by $U$, indicated here by the dotted regions. This is
  achieved if the minima of these standing waves are
  positioned at the auxiliary arrays, leaving them unaffected, and
  their maxima are along the register arrays.}\label{second}
\end{figure}

To perform a one-qubit gate on a certain qubit, $q_1$, let us first move the
pointer next to it, by successive SWAP operations, as presented in the
previous subsection. We then perform, by a Raman transition, the
rotation $U$ on all the register qubits without affecting the
auxiliary qubits. This is possible if we activate the Raman transition
between the ground states $|0\rangle$ and $|1\rangle$ of the atom by
two standing laser fields along the $y$ direction with periodicity $2
\lambda$, as illustrated in Figure \ref{second}. 
Then we perform a control phase gate (CP) between the auxiliary and the
register arrays. This is realised by similar standing laser
fields to those presented in Figure \ref{first}(b). In contrast to the fields
we used to SWAP the qubits, where we lowered the intensities for both
modes, now we aim to activate the
tunnelling of only one of the atomic species \cite{Pachos:2003a} giving,
effectively, a CP gate. This will only apply $\sigma_z$ to $q_1$, as the rest of the
qubits, coupled to the $|0\rangle$ states of the auxiliary qubits, will
not be affected. Next, we apply the inverse
rotation, $U^\dagger$, to all the register qubits using the same laser
configuration as we did for the initial Raman transition. The overall
effect will be that all the register qubits will return to their
original state, while the targeted one will be rotated by
$U\sigma_zU^\dagger$, a one-qubit gate. Note
that for the most general one-qubit gate, we should apply CP twice
to give the evolution $A\sigma_zB\sigma_zC$, where $ABC=\openone$
\cite{nielsen}. However, a universal set of gates can be achieved
without needing to use this more general form.

\subsection{Two-qubit gates with common addressing} 

In order to perform a two-qubit gate between two particular qubits,
$q_1$ and $q_2$, of the register, we move the qubits such that they
neighbour the pointer. We have already specified how to do this if the
qubits are separated by an even number of qubits. If not, then we
have to move the pointer next to $q_1$ (or $q_2$, whichever is more
convenient) and perform a controlled-SWAP, causing the separation
between $q_1$ and $q_2$ to be reduced by 1 site and then they can be
moved together. The need to move the two qubits together is common to all quantum computing schemes that rely on short range interactions and so, in comparison, this procedure does not introduce any additional overhead.
\begin{figure}[!htp]
\begin{center}
\includegraphics[width=12.5cm]{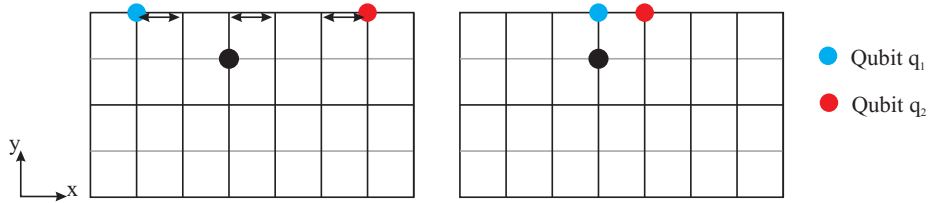}
\end{center}
\vspace{-0.5cm}
\caption[fourth]{The process of bringing $q_1$ and $q_2$ together for
  creating a 2-qubit gate between them.}\label{fourth}
\end{figure}

\begin{figure}[!htp]
\begin{center}
\subfigure[]{\includegraphics[width=10cm]{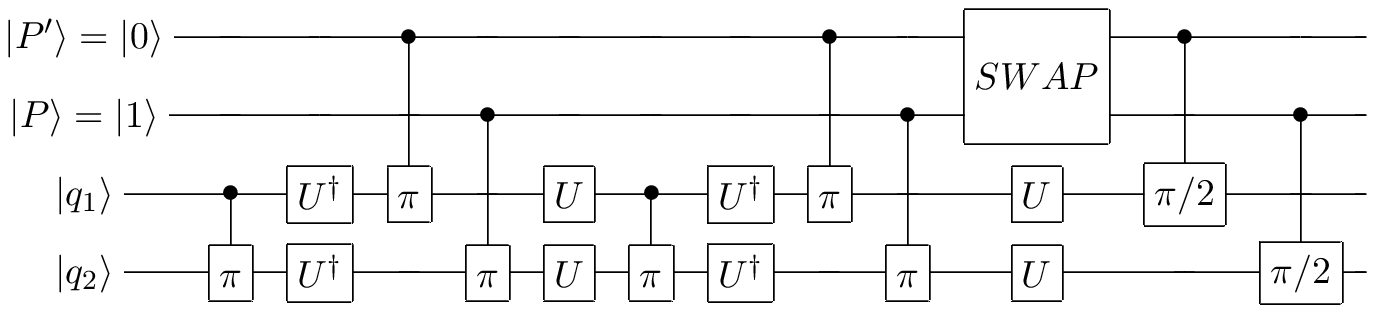}}
\hspace{1cm}
\subtable[]{
\begin{tabular}[b]{|c|c|}
\hline
$H^{\text{c}-\sigma_z}_1$	& $V^\text{c-W}$	\\
$V^\text{c-W}$		& $H^\text{SWAP}_3$	\\
$H^{\text{c}-\sigma_z}_1$	& $V^{\text{c-}\sqrt{\sigma_z}}$	\\
\hline
\end{tabular}
\label{tab:ccnot}
}
\end{center}
\vspace{-0.5cm}
\caption[ccNOT]{The algorithm to perform a controlled-controlled-NOT
around a square is presented. The notation for the program is defined
in Figure \ref{offset}(b). In this notation, the programmed gate of (b) is
written as $C^{1,4}_1$. This circuit is completely general in that it
does not assume any of the bits to be classical. The employed
operations are $U=e^{-i\sigma_x\pi/8}$ and   $W=U^\dagger
\sigma_zU$. Specifically in (a) the algorithm is presented to apply a
two-qubit gate between $q_1$ 
and $q_2$ when $q_1$, $q_2$ and the pointer, $P$, are all on the same
square. $P'$ is the other auxiliary qubit on the square, and starts in
the state $\ket{0}$. In (b) the commands required for performing the
cc-NOT are presented, where the notation is given in Figure
\ref{offset}(b). The commands should be applied from top to bottom within a column and then from left to right.}
\label{ccNOT}
\end{figure}

By performing a three-qubit gate between $q_1$, $q_2$ and the pointer,
such as the controlled-controlled-NOT gate (C$^2$NOT) we effectively obtain a
two-qubit gate between the targeted qubits.
This three qubit gate can be constructed out of globally applied single qubit rotations and
two qubit gates, which we have already used to create a localised single qubit gate. A suitable algorithm is given in Figure
\ref{ccNOT}. We also use this to implement the controlled-SWAP.
This is sufficient to create a 2-qubit gate between $q_1$ and $q_2$
provided they are in the same row (i.e. on the same register). To
perform such a gate between two qubits on different rows, we first
have to move the qubits so that they are within one column of each
other. This may require a controlled-SWAP gate using the pointer. The
pointer then performs a controlled-SWAP to bring $q_1$ onto the
auxiliary array. We then move the auxiliary array such that it is
adjacent to the register containing $q_2$. We are then in a position
to perform our gate, after which $q_1$ should be returned to its
original position. The combination of all these procedures results in
universal quantum computation for the two dimensional register, and
can be trivially extended to three dimensions.

\subsection{Qubit Measurement}

The final step in performing quantum computation is a measurement stage. This can also be accomplished with the help of the pointer. We move the pointer so that it is on the same square, but diagonally opposite, the qubit we wish to measure, $\ket{q_1}=\alpha\ket{0}+\beta\ket{1}$. We then perform the C$^2$NOT procedure presented in the previous subsection, but rotated by $90^{\text{o}}$. The effect of this is to entangle $q_1$ with the auxiliary qubit adjacent to the pointer, giving the state $\alpha\ket{00}+\beta\ket{11}$. If we then measure the auxiliary array using a standing wave with double the lattice period, then this will correctly measure $q_1$, without affecting the register array (except for the one qubit that we measure).

Practically, there is a potential problem in implementing this idea
with global addressing. When we perform a 
measurement as specified in \cite{Beige}, we choose which state to
measure, either $\ket{0}$ or $\ket{1}$. If this state is occupied,
then we are at risk of losing the atom from the optical lattice.
Hence, we can't measure the whole line of auxiliary
qubits because we would lose the pointer. This problem can be avoided if we use a
technique similar to that which we are using on the single-qubit
gates, and only measure every other qubit on the auxiliary array.

\subsection{Pointer Initialisation}

\begin{figure}[!htp]
\begin{center}
\includegraphics[width=13cm]{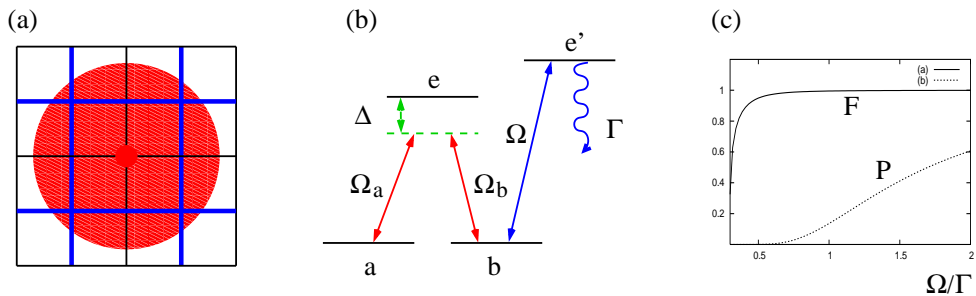}
\end{center}
\vspace{-0.5cm}
\caption[addressing]{The initialisation of the pointer. (a) The laser
profile that aims at the qubit in the centre (pointer). The thicker, blue, lines
indicate the lattice used for measuring state $\ket{1}$ (level $b$)
around the pointer. (b) Energy levels and interactions for
implementing the quantum Zeno effect that keeps level $b$ empty by
strongly coupling it with state $\ket{e^\prime}$. (c)
The fidelity, $F$, (solid line) that the neighbours to the pointer atom
remain in state $\ket{0}$ and the success rate, $P$, (dotted line) for
having no photon emission, with respect to the amplitude
$\Omega$.}\label{addressing}
\end{figure}

Up to now, we have assumed that we can create a pointer on a single
lattice site. Technically, this is not a straightforward procedure,
hence the desire for global addressing. However, as a one-off effect, we
can achieve a single qubit rotation. Recall that, before we create the
pointer, the system consists of a lattice of qubits all in the $\ket{0}$
state. The way in which we intend to create a pointer is to use a
tightly focused laser to create a $\sigma_x$ rotation on a single
qubit. In general, this laser will not have a small enough Gaussian
profile and it will create small rotations on neighbouring qubits. To circumvent this problem, one can
impose a second lattice of
double wavelength along both dimensions, as seen in Figure
\ref{addressing}(a). This lattice should address
the eight nearest neighbours to the qubit we
wish to rotate, causing continuous measurements on their state
$\ket{1}$ and thus preventing population of this state through
the quantum Zeno effect \cite{Beige,Beige1}. This is accomplished by applying an additional laser with amplitude $\Omega$ that
couples the state $\ket{1}$ to an excited state $\ket{e^\prime}$
which spontaneously emits photons at a rate $\Gamma$. This is
illustrated in Figure \ref{addressing}(b), where the unwanted Raman
transition between $a$ and $b$ is also
depicted. The efficiency of this scheme has been studied by a
simulation of the evolution of the state of the atoms that neighbour
the pointer.
Figure \ref{addressing}(c) shows the results of this simulation,
plotting the fidelity of keeping the population of the
neighbours fixed in the $\ket{0}$ state, $F$, and the success rate, $P$. The
success rate is the probability that no photon emission occurs during
the process. We have assumed that the laser has a width of $0.8
\lambda$. This
proves that significant suppression takes place for
physically sensible laser profiles, since the fidelity can be brought
close to 1. If photons are emitted, then
they can be measured and the initialisation process can be
repeated. In principle, further lattices can be 
applied, if necessary, to suppress the effect on the next-nearest
neighbours.

Vollbrecht et al., in \cite{Vollbrecht}, introduce a potentially very
interesting concept which generates pointer qubits from imperfections
in the lattice (the globally addressed
computation component of their paper is only of secondary
nature). This idea not only creates
pointers, but also removes any imperfections in the loading of atoms
into the lattice. We can, in fact, incorporate the idea into our
scheme as an alternative procedure for preparing the pointer. We would
do this by following the scheme presented in
\cite{Vollbrecht} until ready to perform the computation. At this
point, we add two additional steps. These involve, firstly, performing
a controlled-NOT operation as defined in their scheme, so that the qubit adjacent to the pointer is placed
in the $\ket{1}$ state. Secondly, we reject the pointer qubit, so that
there is only one qubit in each lattice site, hence we have
converted from their pointer into our pointer. However, the authors of
this paper accept that the number of steps required for this
preparation is currently very demanding from an experimental
perspective. It also requires a different set of controls to those
which we use in the rest of the computation. Both of these factors
make it preferable to use quasi-single qubit addressing outlined
above.  

\section{Physical implementation of superlattices}

\begin{figure}[!htp]
\begin{center}
\includegraphics[width=14cm]{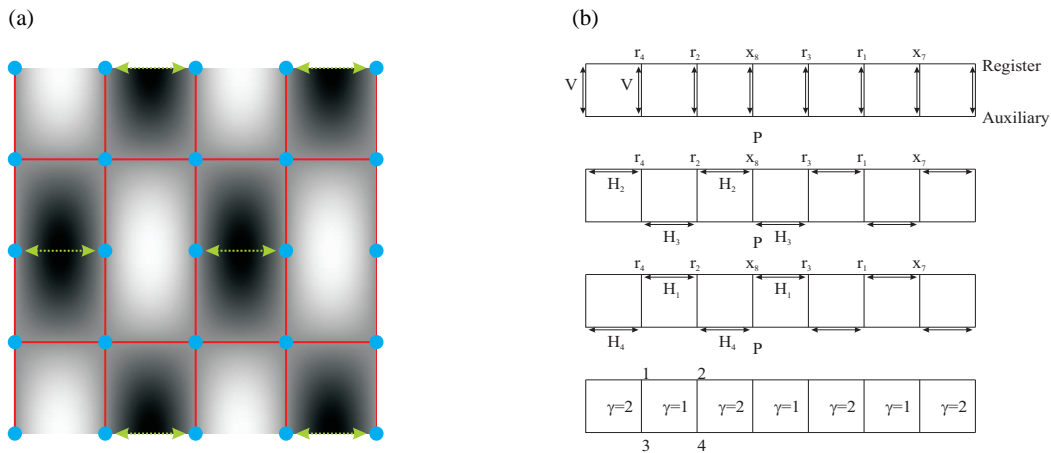} 
\end{center}
\caption[offset]{(a) Potential offset from trapping potential required for
  generating alternate couplings. The red lines indicate lines of no
  offset. The qubits are indicated by blue circles. The white
  and black regions show positive and negative offsets
  respectively. The arrows indicate which qubits are
  interacting. 

(b) Definition of the initial labelling of qubits and
  labels for the different interactions. $V^\text{c-NOT}$ means apply
  a c-NOT operation vertically, with the auxiliary as
  control. $C^{q_2,P'}_{\gamma}$ means perform a cc-NOT operation with
  qubit $q_2$ the target, and $P'$ the unused qubit on the square,
  $\gamma$.}\label{offset}
\end{figure}

In order to perform the previously presented quantum gates
we need to enable couplings on alternate rows and between
alternate pairs. The required arrangement of lasers can be calculated
by specifying the potential offset from the base trapping potential
that we need to create across the 2D structure. Specifically, we need
to ensure that the zero offsets appear exactly where we want no
interactions to occur. For example, let us assume that 
we want to create interactions as shown in Figure \ref{offset}(b), and
that the vertices where the qubits are located are separated by a
distance of $\lambda/2$. The potential offset, $V_{\text{off}}$, that we
require can thus be written as
\begin{equation}
V_{\text{off}}=\sin\left(\frac{2\pi
  x}{\lambda}\right)\cos\left(\frac{\pi y}{\lambda}\right)
\end{equation}

This expression can be expanded to give a sum of sine terms. Each term
has a period and a direction. The period, $d$, specifies the angle
between the pair of lasers that is required to create that term and is
given by
$$
\sin\left(\frac{\theta}{2}\right)=\frac{\lambda}{2d}
$$
This potential, illustrated in Figure
\ref{offset}(a), can be implemented by the combination of two
independent pairs of lasers, each one producing a potential offset,
$V^i_\text{off}$, given by
\begin{eqnarray}
V_{\text{off}}=\sum_{i=1}^2 V^i_{\text{off}}=&&
{1 \over 2}
\sin\left[\frac{2\pi}{\lambda} \left(x-\frac{y}{2}\right)\right]+ 
{1 \over 2}\sin\left[\frac{2\pi}{\lambda}
  \left(x+\frac{y}{2}\right)\right] 
\end{eqnarray}
For example, taking the term
$V^1_\text{off}$, the required laser field has a wave vector along the
direction of $(2\vec{i} -\vec{j}) /\sqrt{5}$ and a
period $d=2\pi/(\sqrt{5}\pi/\lambda)=2\lambda/\sqrt{5}$. 
As a result, we can employ lasers of the same wavelength as those creating the trapping potential, where the
doubling of the wavelength is produced by having $\theta\approx
68.0^{\text{o}}$. Similar setups can be used for generating the other
control procedures of the previous section.

%\begin{figure}[!htp]
%\begin{center}
%\includegraphics[width=8cm]{define_15.eps}
%\end{center}
%\caption[define_15]{Definition of the initial labelling of qubits and
%  labels for the different interactions. $V^\text{c-NOT}$ means apply
%  a c-NOT operation vertically, with the auxiliary as
%  control. $C^{q_2,P'}_{\gamma}$ means perform a cc-NOT operation with
%  qubit $q_2$ the target, and $P'$ the unused qubit on the square,
%  $\gamma$.}\label{define_15}
%\end{figure}

As a final point we would like to consider the influence of the superlattices on the
harmonicity of the trapping potentials of the atoms. Around the
potential minima, the superposition effect can be given
by expansions of sine or cosine functions. The qubits that
remain uncoupled get only even powers in the expansion including
quadratic terms, and hence the location of the qubits remain
unchanged. However, the qubits that become coupled obtain an $x$ term,
and hence the trapping minima for the coupled qubits actually move
together. This is not a problem provided the offset potential remains
small compared to the trapping potential, and the superlattices are
turned on adiabatically so the atoms remain in their ground states of
the trapping potential. This also demonstrates that the trapping
frequencies of the interacting qubits
remain unchanged when the additional lattice field is introduced.

\section{Proposal for the experimental realisation of factoring 15}

The standard experimental demonstration of a quantum computation is to
factor 15 \cite{nmr_guys}, the smallest meaningful factorisation,
since the method fails for even numbers and powers of primes. The
employed algorithms include a significant element of simplification
(for a full description of such methods, see \cite{nmr_guys,
Beckman:1996}). Since we can already factor 15 by classically
calculating all the steps in the algorithm, we can quickly realise
that most of the work is redundant, enabling us to reduce the required
control steps. This won't be possible when we consider factoring
much larger numbers.

The general factoring scheme, for a number $N$, which is $K$ bits
long, consists of taking $2K$ qubits (the first register), and
applying the Hadamard to each of the qubits. This creates an equally
weighted superposition of all numbers, $x$, from 0 to $2^{2K}-1$. We
then take another $K$ qubits (the second register) and calculate
$a^x\text{ mod }N$, thus entangling them with the first register.
$a$ is a number that is randomly selected from the numbers
smaller than $N$ which satisfies $\text{gcd}(a,N)=1$. So, for $N=15$,
$a\in\{1,2,4,7,8,11,13,14\}$. This calculation will, in general, also
require some ancillas to act as scratch space for the
calculation. The next step is to apply the inverse Fourier
transform on the first register and then measure these qubits. The
result is used as the input to a classical continued fractions
algorithm, which will finally yield one of the factors of $N$. 

We start the simplification process by noting that, in the case of
$N=15$, $a^4\text{ mod }15=1$ for all $a$. This means that only the 2
least significant bits of $x$ affect the calculation on the second
register. Furthermore, if we `randomly' select $a=11$ (say), we find
that $a^2\text{ mod }15=1$ and hence only the least significant bit of
the first register matters. The computation that we have to perform
then becomes very simple, while still creating the state 
$$
\sum_{x=0}^{2K-1}\ket{x}\ket{a^x\text{ mod }N}.
$$
\begin{figure}[!htp]
\begin{center}
\subfigure[]{
\includegraphics[width=0.4\textwidth]{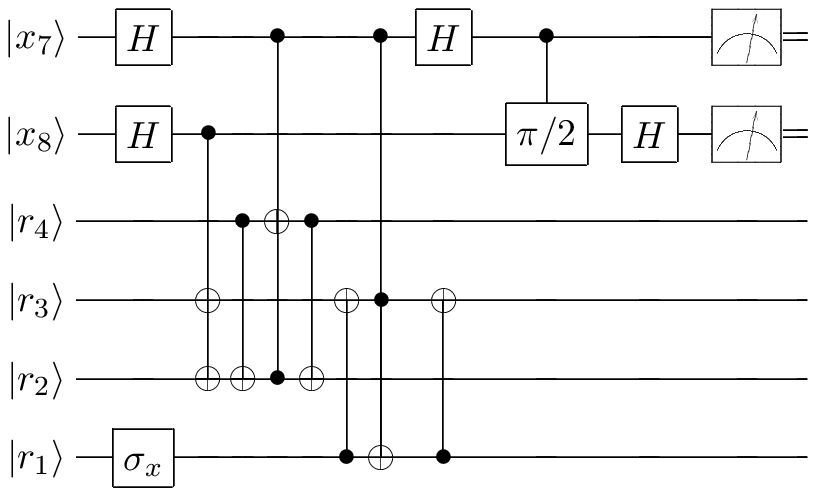}
}
\hspace{1cm}
\subtable[]{
\begin{tabular}[b]{|c|c|c|c|}
\hline
$V^\text{c-H}$		& $H^\text{SWAP}_4$	& $C^{2,4}_2$		& $V^\text{c-U}$	\\
$C^{1,3}_2$			& $C^{3,2}_2$		& $H^\text{SWAP}_4$	& $C^{1,4}_2$		\\
$C^{2,4}_1$			& $H^\text{SWAP}_3$	& $H^\text{SWAP}_3$	& $V^{\text{c-U}^{\dagger}\text{H}}$	\\
$H^\text{SWAP}_3$	& $H^\text{SWAP}_4$	& $C^{3,2}_2$		& $H^\text{SWAP}_3$	\\
$V^{\text{c}-\sigma_x}$		& $H^\text{SWAP}_3$	& $V^\text{c-H}$	& $H^\text{SWAP}_4$	\\
$H^\text{SWAP}_4$	& $H^\text{SWAP}_4$	& $H^\text{SWAP}_3$	& $C^{1,3}_1$		\\
$V^{\text{c}-\sigma_x}$		& $C^{1,3}_1$		& $C^{1,3}_2$		& \\
$H^\text{SWAP}_3$	& $H^\text{SWAP}_4$	& $H^\text{SWAP}_4$	& \\ 
$V^\text{c-H}$		& $H^\text{SWAP}_3$	& $H^\text{SWAP}_1$	& \\
\hline
\end{tabular}
}
\end{center}
\vspace{-0.5cm}
\caption{The algorithm required for the `hard' case of factoring
  15. (a) shows the circuit diagram, and (b) provides the required set
  of commands. The commands should be applied from top to bottom within a column and then from left to right. The notation is defined in Figure
  \ref{offset}(b). Here $U\sigma_xU^\dagger=\sqrt{\sigma_z}$.} 
\label{factoring}
\end{figure}
The more standard case to demonstrate is the choice of $a=7$, where we
have to act on 2 of the bits of $x$. The circuit diagram for this is
given in Figure \ref{factoring} \cite{nmr_guys}. In Table
\ref{factoring}(b) we give the set of commands required to perform
this factorisation, with the notation being defined in Figure
\ref{offset}(b).

The minimum device size is a grid of $18\times 2$ qubits. In general, the size of the optical lattice for computing on $N$ qubits needs to be $3N\times2$ qubits. The need for this can be seen in Figure \ref{offset}(b) since if we were to apply $H_2^{\text{SWAP}}$, qubits $r_4$ and $x_7$ would try to move into empty lattice sites if the device size was just $N\times 2$, but the interaction has different effects if those lattice sites are empty.

\section{Error avoidance and error correction}

\subsection{The Auxiliary Qubits}

As with any proposal for quantum computation, decoherence is a
significant issue that cannot be neglected. In a global control
scheme, its significance only gets amplified. Such a scheme
necessarily introduces more computational steps and so the algorithm
will take longer to run, thus increasing the build-up of errors. Even
more demanding is the requirement that
our pointer qubit and, in fact, all the other qubits in the auxiliary
array, are error free. If an error occurs in the auxiliary array, it
will affect every gate throughout the rest of the computation.

Such an error would be catastrophic for our computation, and needs to
be prevented. This can be achieved by noting that all the
qubits of the auxiliary array are in classical states. These
classical states are eigenstates of $\sigma_z$, and are thus
unaffected by phase-flip errors. The effect of bit-flip errors can be
reduced by using the quantum Zeno effect \cite{Beige}. In principle,
by continuous measurement, the probability of a bit-flip can be
reduced to zero. Since all errors can be described in terms of phase
flips, bit flips or a combination of the two, this renders the
auxiliary array error free.

While applying the quantum Zeno effect we do not wish to lose the
pointer during the measuring procedure. Consequently, we have to
employ the same optical lattice that
only measures every other qubit on the auxiliary array. This also means
that we can never measure the state of the pointer. We have to use the
measurement result of the other qubits to indicate how likely it is
that an error has occurred on the pointer. If some of the auxiliary
qubits are found in the $\ket{1}$ state, then we have to consider it
likely that the pointer has also been affected and stop the
computation.
In the following subsection, we present how error correction can be
performed on the register qubits. In principle, similar concepts can
be applied to the pointer. However, our current method requires
significant modification of the physical models, and is in need of
optimisation. This is an avenue for future study.

\subsection{The Register Qubits}

\begin{figure}[!htp]
  \begin{center}
    \leavevmode
\resizebox{10cm}{!}{\includegraphics{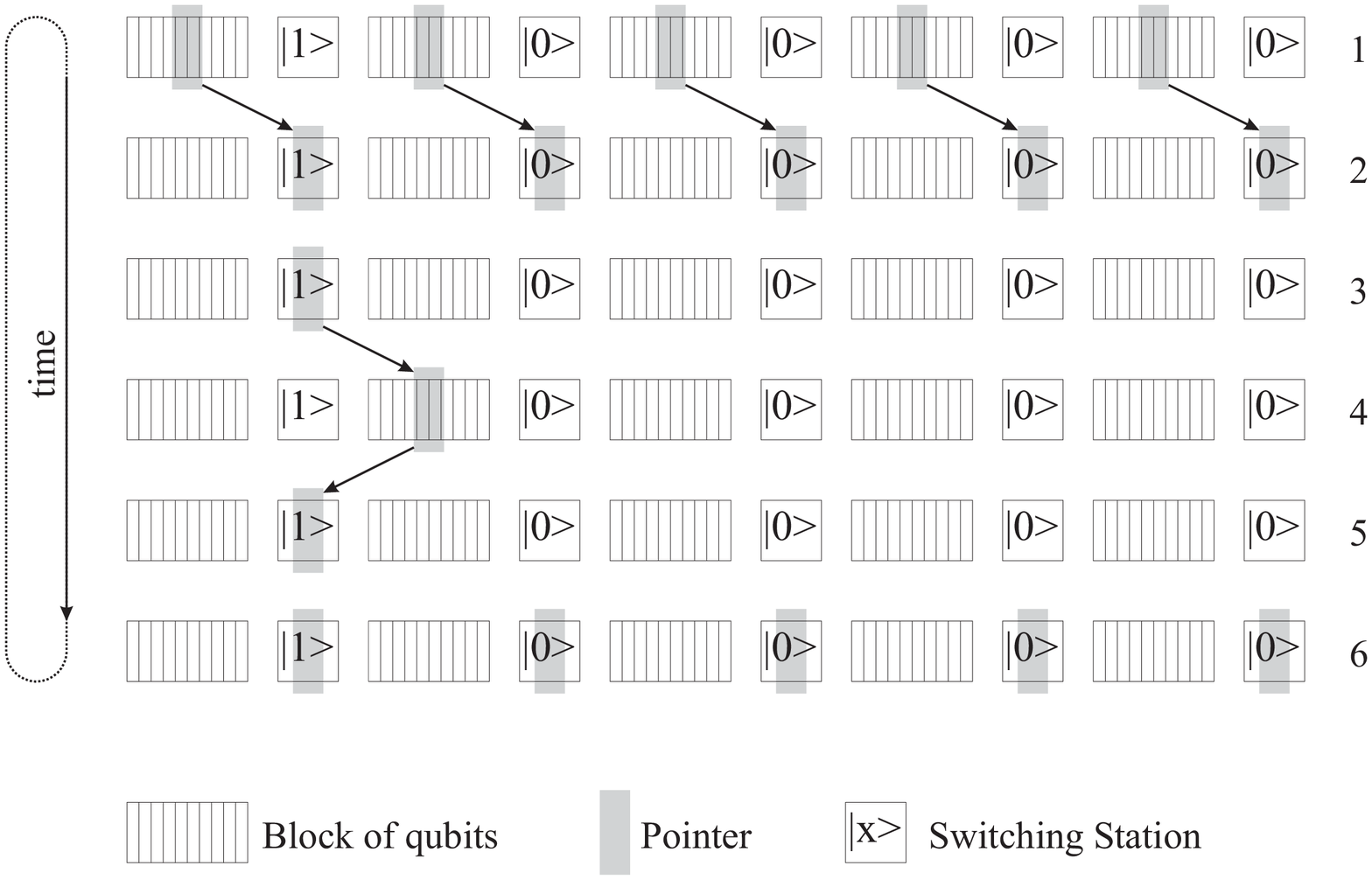}}
\end{center}
\caption{The process of alternating between error correcting and
  algorithmic phases of computation.
(1) There is a pointer for every block of qubits. We send out pulses
that cause a pointer to perform error correction on a single qubit
block. 
(2) Pointers move onto the switching stations.
(3) All but one of the pointers are deactivated.
(4) Remaining pointer performs steps in the algorithm until another
error correction phase is required. 
(5) Pointer moves back onto switching station.
(6) All switching stations get reactivated.
} 
\label{figure:process}
\end{figure}

Recent work \cite{Benjamin:2003b} has shown that error correction can
be implemented on globally controlled quantum computers, and that the
architecture even supports fault tolerance. The basic idea is that the
qubits are divided up into blocks of $m$
qubits. These $m$ qubits will constitute one encoded qubit for error
correction. A typical error correction phase of a computation
involves extracting a syndrome measurement from the encoded qubits,
thus finding out what errors have occurred and, depending on the
results held on ancillas, correcting for the error. The way one can
implement this in a globally controlled structure is to add 
extra quantum gates that feed back from the ancillas to correct for
the errors, without ever actually measuring them. We also require a
`switching station', which is a classical pattern of qubits, for every
encoded block of qubits. This switching station allows us to switch on
and off pointers with the correct sequence of global pulses. The
patterning required for error correction is
simply a $\ket{1}$ for one of the blocks, and a $\ket{0}$ for
all the others, as in Figure \ref{figure:process}. 
If we start with a
pointer every $m$ qubits (and we 
will select $m$ to be even for the sake of the quantum Zeno effect, as
described above), then the pointers start off with the correct
parallelism for error correction. To switch to a phase where we want
to apply operations to a single qubit, we just perform a
controlled-operation using the switching station as the control. 

There are several drawbacks with this implementation of error
correction. The first is that such algorithms are currently outside
experimental feasibility because they require several thousand steps
(\cite{Benjamin:2003b} gives specifications for two common codes),
which would require a longer time to perform than current decoherence
times. The second is that these switching stations are susceptible to
errors. We will have similar problems applying the quantum Zeno effect
to them as we do to the auxiliary qubits. Finally, we have to take
care of the ancillas. At the start of each error
correcting phase, we require our ancillas to be in the $\ket{0}$
state. At the end of this phase, they will contain an unknown error
syndrome. These qubits either have to be reset, or replaced by
fresh ancillas. In the discussion of 3D structure in the following
section, we will present how the third dimension could be used to
provide a significant supply of fresh ancillas. This greatly
simplifies the experimental implementation of quantum error correction
algorithms.

\section{Proposed computational schemes and conclusions}

There are many different ways that our scheme can be used for
performing quantum computation with three dimensional optical
lattices. The choice of which one should be used can 
be determined by setting different priorities, such as efficiency
of scaling, signal sizes etc. We outline some of the possible uses below.

The conceptually simplest model is to arrange the computation on a one
dimensional ladder, consisting of $N$ qubits on a single register array
and a single auxiliary array, also consisting of $N$ qubits. This could be
repeated across the whole three dimensional structure, so we would
have approximately $N^2$ identical copies, all running in
parallel. This is the least scalable architecture, since qubits are
separated by $O(N)$ steps. However, it is very useful in terms of
signal strength
when we make a measurement at the end of the computation. Using the
computer with many copies running in parallel gives an
ensemble computation, where expectation values of the computational
result can be given at the end.
%Noted that this gives ensemble computation
Alternatively, we can perform computation on a two dimensional grid by
moving the registers relative to the auxiliaries with a series of SWAPs
in the $y$ direction. All qubits are within
$2\sqrt{N}$ steps of each other, and so the structure scales
more easily. We could also get a reasonable signal strength due to the
parallel planes in the third dimension.
As a final alternative, we could perform computation on a single
plane, leaving all the other planes in the $\ket{0}$ state, kept
there possibly by the quantum Zeno effect. These other planes can 
then be used for ancillas in the error correction phase. They can be
easily accessed as the computational plane can be moved through the
other planes in the same way that the auxiliary array can be moved
past the register arrays.

Naturally, other possible arrangements exist, but these three
illustrate some of the simple ideas that can be used.
It is also important to remember that we are free to rearrange the
labelling of our qubits so as to minimise the path of the
pointer. Some operations can also be reorganised to minimise its
path. These procedures can have a significant effect on the number of
steps required to implement an algorithm, and hence also on the
errors, and the practicality. Such an example has been given in a
previous section where a convenient arrangement was given for the
implementation of factoring 15. The significant remaining issue
is the number of gates that can be implemented within the system's
decoherence time.

%compare and contrast this scheme to those of Zoller:1 and Vollbrecht. Both referees wanted this
%also deals with point 3 of the first referee, alluded to by sewcond referee.
Our scheme has significant differences in comparison to those
already proposed \cite{Zoller:1, Vollbrecht}. At the most fundamental
level, we interact qubits in a different way, making use of the
tunnelling interaction \cite{Pachos:2003a} instead of collisional
couplings \cite{Greiner:2003a}. These are just two different
experimental techniques and the current state of the art provides little
to pick between them for performing global addressing quantum
computation. The collisional schemes, however, have
significant complications and/or drawbacks. Both these schemes use an
additional qubit whose presence, or lack thereof, is crucial in
performing quantum gates. This gives the pointer a very special
position. In the present
proposal, the pointer is not that special, it is just the same as
every other qubit, experiencing the same fields etc. This makes
experimental realisation easier, but also contributes to the
complexity of the ideas required for the scheme.
The work of \cite{Zoller:1} has significant issues associated with the
initialisation procedure. They require a filling of one qubit on every
other lattice site, which has been experimentally demonstrated in
\cite{Peil}. However, the pointer (referred to as a marker atom) is
an extra atom that has to be inserted into one of the gaps.

Our proposal extends the work presented in these papers by including
control sequences that are particularly suitable for global addressing
and it incorporates the concepts of error correction and avoidance
\cite{Benjamin:2003b}. It is unclear as to whether the scheme in
\cite{Vollbrecht} could be generalised to give the regular patterns
required for error correction and fault tolerance, whereas the
pointers in our proposal have no special place and can easily be
prepared once a single pointer has been initialised.

\acknowledgments

This work was
supported by EPSRC and a Royal Society URF.

\bibliography{references}

\begin{thebibliography}{25}
\bibitem{comment} We shall later demonstrate how this can be done in practice.
\end{thebibliography}

\end{document}